\documentclass[amssymb,prl,twocolumn,longbibliography,notitlepage,superscriptaddress]{revtex4}
\usepackage{amsmath,amssymb,mathrsfs,bm}
\usepackage{graphicx,dcolumn,times}
\usepackage{subfigure}
\usepackage{color}
\definecolor{blueblack}{rgb}{0,0,0.6}
\usepackage{hyperref}
\hypersetup{hypertex=false,colorlinks=true,linkcolor=blueblack,anchorcolor=blueblack,citecolor=blueblack,urlcolor=black}

\usepackage{soul}

\newcommand{\Zcite}[1]{\textcolor{red}{[cite]}}

\begin{document}
\title{Quantum interference between non-identical single particles}

\author{Keyu Su}
\affiliation{Guangdong Provincial Key Laboratory of Quantum Engineering and Quantum Materials, School of Physics and Telecommunication Engineering, South China Normal University, Guangzhou 510006, China}

\author{Yi Zhong}
\affiliation{Guangdong Provincial Key Laboratory of Quantum Engineering and Quantum Materials, School of Physics and Telecommunication Engineering, South China Normal University, Guangzhou 510006, China}

\author{Shanchao Zhang}
\affiliation{Guangdong Provincial Key Laboratory of Quantum Engineering and Quantum Materials, School of Physics and Telecommunication Engineering, South China Normal University, Guangzhou 510006, China}
\affiliation{Guangdong-Hong Kong Joint Laboratory of Quantum Matter, Frontier Research Institute for Physics, South China Normal University, Guangzhou 510006, China}

\author{Jianfeng Li}
\affiliation{Guangdong Provincial Key Laboratory of Quantum Engineering and Quantum Materials, School of Physics and Telecommunication Engineering, South China Normal University, Guangzhou 510006, China}
\affiliation{Guangdong-Hong Kong Joint Laboratory of Quantum Matter, Frontier Research Institute for Physics, South China Normal University, Guangzhou 510006, China}

\author{Chang-Ling Zou}
\affiliation{CAS Key Laboratory of Quantum Information,  University of Science and Technology of China, Hefei 230026, China}

\author{Yunfei Wang}
\email{yunfeiwang2014@126.com}
\affiliation{Guangdong Provincial Key Laboratory of Quantum Engineering and Quantum Materials, School of Physics and Telecommunication Engineering, South China Normal University, Guangzhou 510006, China}
\affiliation{Guangdong-Hong Kong Joint Laboratory of Quantum Matter, Frontier Research Institute for Physics, South China Normal University, Guangzhou 510006, China}
\author{Hui Yan}
\email{yanhui@scnu.edu.cn}
\affiliation{Guangdong Provincial Key Laboratory of Quantum Engineering and Quantum Materials, School of Physics and Telecommunication Engineering, South China Normal University, Guangzhou 510006, China}
\affiliation{Guangdong-Hong Kong Joint Laboratory of Quantum Matter, Frontier Research Institute for Physics, South China Normal University, Guangzhou 510006, China}
\affiliation{Guangdong Provincial Engineering Technology Research Center for Quantum Precision Measurement, South China Normal University, Guangzhou 510006, China}

\author{Shi-Liang Zhu}
\email{slzhu@scnu.edu.cn}
\affiliation{Guangdong Provincial Key Laboratory of Quantum Engineering and Quantum Materials, School of Physics and Telecommunication Engineering, South China Normal University, Guangzhou 510006, China}
\affiliation{Guangdong-Hong Kong Joint Laboratory of Quantum Matter, Frontier Research Institute for Physics, South China Normal University, Guangzhou 510006, China}

\date{\today}

\begin{abstract}
 Quantum interference between identical single particles reveals the intrinsic quantum statistic nature of particles, which  could not be interpreted through classical physics.  Here, we demonstrate quantum interference between non-identical bosons using a generalized beam splitter based on a quantum memory. The Hong-Ou-Mandel type interference between single photons and single magnons with high visibility is demonstrated, and the cross-over from the bosonic to fermionic quantum statistics is observed by tuning the beam splitter to be non-Hermitian. Moreover, multi-particle interference that simulates the behavior of three fermions by three input photons is realized.  Our work extends the understanding of the quantum interference effects and demonstrates a versatile experimental platform for studying and engineering quantum statistics of particles.
\end{abstract}
\maketitle

Multi-particle quantum interference, such as the well-known Hong-Ou-Mandel (HOM) interference, reveals the quantum statistic nature of particles~\cite{HOM}. There is great research interest in studying the distinct physics of the bosonic and fermionic quantum interference effects~\cite{PQS2012,BSamp2013NP,antiHOM,HOMNJP2012}. The beam splitter (BS) is a fundamental element in conducting quantum interference research. It creates the superposition of particles in different output ports and realizes the interference of amplitudes for particles in each port. Conventionally, the linear BS that separates particles into different spatial modes has been used to demonstrate  quantum interference between photons and photons, magnons and magnons, plasmons and plasmons, and even between massive particles (trapped atoms) ~\cite{HOMreview2021,twoexcitations,TwoplasmonHOM2014,plasmonHOMcirciut2013,AtomicHOM2015,CindyRegal2014Science}. These demonstrations promise a wide range of potential single-particle level quantum devices for future applications. For instance,  quantum interference between single photons provides the single-photon nonlinearity for multi-qubit quantum gates~\cite{LQC2001,LQCRev2007,PNAS2011} and thus lies at the heart of quantum information processing and quantum communications~\cite{PanRMP2012,DLCZ2001,QRepeaterRMP2011}.

Although most of the previous experimental progress has been achieved with identical particles using unitary BS, the principle of quantum interference is not limited to identical particles or Hermitian particle interactions, provided that the coherent superposition between single particles could be realized. Therefore, extending quantum interference to a generalized BS is of fundamental importance.  HOM interference between optical photons with different colors has  recently been demonstrated with the assistance of a coherent frequency converter~\cite{twocolorNP}. HOM interference between a single magnon excitation and a photonic coherent state has also been demonstrated\cite{chenjf2022prl}. In addition,  fermion-like quantum interference using a non-Hermitian BS has been observed with identical single photons~\cite{NPmetasurface2021,tailorloss2022,lossybs2017} and coherent absorption of the NOON state has been realized~\cite{NOON2016}.  However, for non-identical particles, it is experimentally challenging to prepare single-particle quantum input for each port and to make them indistinguishable in a beam splitter. Therefore, the quantum interference between genuine non-identical single particles remains elusive.

In this Letter, the quantum interference between distinct single particles, i.e., single photons and single magnons, is demonstrated for the first time. We construct a hybrid BS for non-identical particles  by realizing  coherent quantum conversion between the stored magnon excitation and photons via indistinguishable dark-state polaritons in a cold atom quantum memory.  In contrast to the conventional linear BS, our hybrid BS for non-identical particles  could be either Hermitian or non-Hermitian, controllable by an external control laser field. Therefore, our versatile hybrid BS demonstrates the cross-over from bosonic bunching statistics (the second-order cross-correlation function $g^{(2)}(0)=0.40$) to fermionic anti-bunching statistics ($g^{(2)}(0)=1.71$), even though the input non-identical particles are bosons. Furthermore,  quantum interference can be extended to more than two particles, with three single photons  engineered to behave as three fermions by non-Hermitian hybrid BSs. Our work unambiguously demonstrates  tunable quantum interference between non-identical single bosons, which provides a new tool for engineering quantum states for hybrid quantum systems.

\begin{figure*}
\centering\includegraphics[width=17cm]{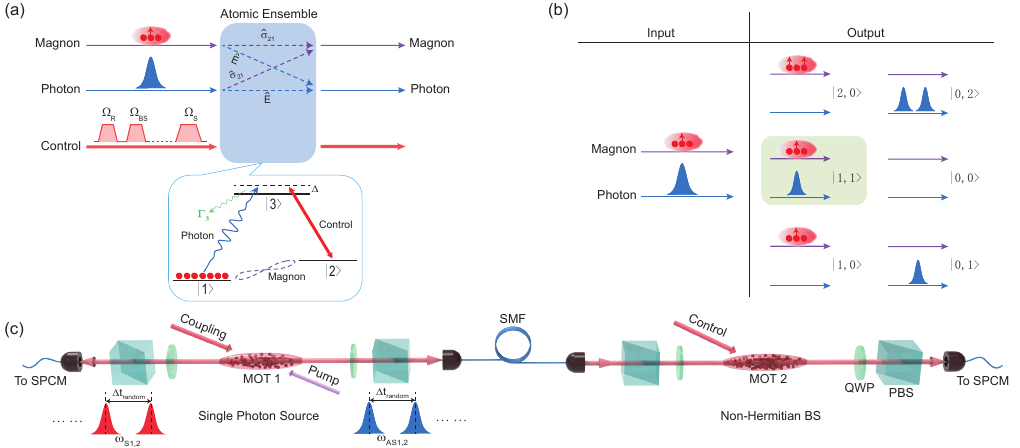}\\
\caption{\label{Fig1}Theoretical and experimental schemes. (a) The theoretical scheme of the magnon-photon HOM interferometer. The magnon ($\hat{\sigma}$) and photon ($\hat{E}$) are both in the form of a dark-state polariton (DSP) in an electromagnetically induced transparency (EIT) medium. The red line above the control laser shows the experimental timing sequences: storage of a magnon ($\Omega _S$), interference between two DSPs ($\Omega _{BS}$) and reading out of the magnon ($\Omega _R$). The insert is a $\Lambda$-type EIT energy diagram: $|1\rangle  = |5{S_{1/2}},F = 2,{m_F} = 2\rangle$,$|2\rangle  = |5{S_{1/2}},F = 3,{m_F} = 2\rangle$,$|3\rangle  = |5{P_{1/2}},F = 3,{m_F} = 3\rangle$, $\Gamma_{3}$ is the spontaneous decay rate of $|3\rangle$, and $\Delta$ is  single photon detuning. (b) The input and output states of the magnon-photon HOM interferometer. (c) Experimental setup. MOT$_{1}$ is a single photon source. The detection of a Stokes photon ($\omega_{Si}$) heralded the generation of an anti-Stokes photon ($\omega_{ASi}$). MOT$_{2}$ is the non-Hermitian beam splitter. PBS:polarization beam splitter, QWP:quarter wave plate, SPCM: single photon counting module, SMF: single mode fiber.}
\end{figure*}

Figure~\ref{Fig1} schematically illustrates the principle and experimental setup for demonstrating  quantum interference between single photons and single magnons. In a cold atom ensemble [Fig.~\ref{Fig1}(a)],  quantum memory could be realized by using an electromagnetically induced transparency (EIT) scheme, by which a control laser could stimulate the coherent conversion between a single flying carrier (photon) and a single collective atomic excitation (magnon)~\cite{HarrisEIT1991,FirstEITQM2001Nature,FirstEITQM2001PRL,RMP2003,QMDSP2002PRA}. Through such a process, the superposition of the photon and magnon could be realized, and  quantum interference between these two distinct bosons becomes possible.

Inside the EIT medium, the hybrid superposition state is essentially a dark-state polariton (DSP)~\cite{QMDSP2002PRA,DSP2000}. When a photon enters the EIT media, a supposition of photon state $\hat E\left( {z,t} \right)$ and excited magnon state $\hat \sigma _{12}\left( {z,t} \right)$ would be generated \cite{DSP2000}:
\begin{equation}
\Psi \left( {z,t} \right) = \cos \theta \hat E\left( {z,t} \right) - \sin \theta \sqrt N {\hat \sigma _{12}}\left( {z,t} \right),
\end{equation}
where $\cos \theta=\Omega _c/\sqrt{\Omega _c^{2}+ g^{2}N}$, $\Omega _c$ is the Rabi frequency of the control laser, $g$ is the atom-field coupling constant, and $N$ is the number of atoms. When the control laser is on, the DSP will propagate in the medium with a slow group velocity and eventually be converted back to photons when leaving the medium. Before leaving the medium, a DSP can be converted to a pure magnon by switching off the control laser adiabatically. Therefore, by controlling the switching timing of the control laser, a hybrid BS with a temporal photonic input and a stationary magnonic input can be represented as 
\begin{equation}
\left[ {\begin{array}{*{20}{c}}
{{M_{out}}}\\
{{A_{out}}}
\end{array}} \right] = \left[ {\begin{array}{*{20}{c}}
{t_{1}}&{r_{2}}\\
{r_{1}}&{t_{2}}
\end{array}} \right]\left[ {\begin{array}{*{20}{c}}
{{M_{in}}}\\
{{A_{in}}}
\end{array}} \right].
\end{equation}
Here, $M_{out}$ ($M_{in}$) and $A_{out}$ ($A_{in}$) denote the magnon and photon states of output (input), respectively. $t_{1}=|t_{1}|e^{i\phi_{1t}}$ ($r_{1}=|r_{1}|e^{i\phi_{1r}}$) represents the transmission (reflection) coefficient of the BS for  input from the magnonic port. $t_{2}=|t_{2}|e^{i\phi_{2t}}$ ($r_{2}=|r_{2}|e^{i\phi_{2r}}$) refers to the transmission (reflection) coefficient of the BS for input from the photonic port.

By manipulating the Rabi frequency of the control laser and optical depth (OD), the loss of the dark-state polaritons induced by atomic spontaneous emission is controllable, and eventually contributes to a tunable non-Hermitian photon-magnon interaction. Consequently, a tunable non-Hermitian beam splitter (NHBS) for these non-identical bosons can be realized, with the non-Hermitian effect being captured by an overall phase difference $\phi _{rt}=\phi_{1r}-\phi_{2t}+\phi_{2r}-\phi_{1t}$ between the reflection and the transmission channel~\cite{chenjf2019}. For a conventional Hermitian and symmetric BS with $|t_{1,2}|^{2}+|r_{1,2}|^{2}=1$, we have $\phi _{rt}=\pi$.  If $|t_{1,2}|^{2}+|r_{1,2}|^{2}<1$, it becomes a non-unitary BS. For the non-unitary BS, the phase difference $\phi _{rt}$ can be obtained as (Sec. II of the Supplemental Materials~\cite{SM}):
\begin{equation}
{\phi _{rt}} = \arg \left[ {1 - {1 \mathord{\left/
 {\vphantom {1 \xi }} \right.
 \kern-\nulldelimiterspace} \xi }} \right] + \arg \left[ {\frac{{\eta \left( {\xi  - 1} \right)}}{{\zeta  - \eta \left( {1 - \xi } \right)}}} \right],
\end{equation}
where $\xi=e^{\frac{-|\Omega_{c}|^{2}}{4(\gamma_{31}-i\Delta)}\tau_{p}}$ and $\zeta=\frac{1}{4}|\Omega_{c}|^{2}\tau_{p}/\gamma_{31}$, $\Delta$ is the single photon detuning of the control laser, $\gamma_{31}$ is the dephasing rate, $\tau_{p}$ is the temporal length of the input single photon, and $\eta$ is the OD of the atomic ensemble. As shown in Fig. \ref{Fig1}(b), there are six possible output states after the magnon-photon HOM interferometer (Sec. III of the Supplemental Material \cite{SM}). Different from the Hermitian BS, with NHBS, two bosonic particles quantum interference  can have an output state of $|1,1\rangle$, where bosons behave like fermions.

\begin{figure*}
\centering\includegraphics[width=17.5cm]{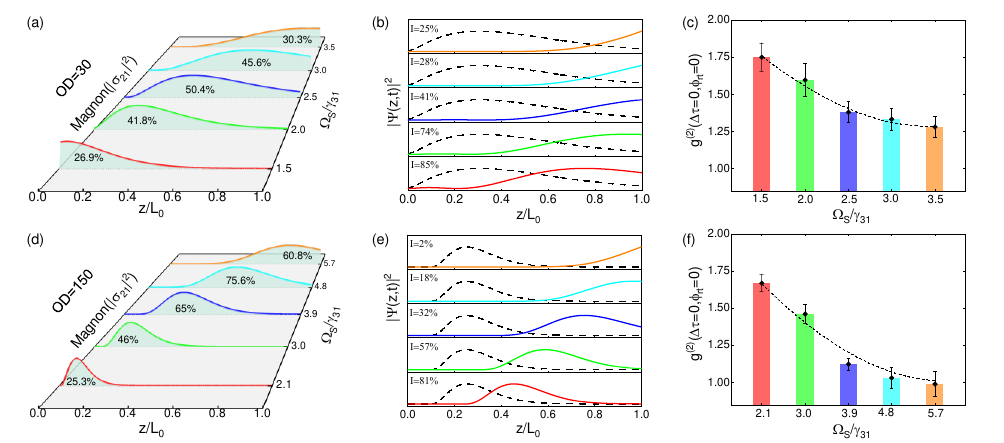}\\
\caption{\label{Fig2} The interference contrast of the magnon-photon HOM interferometer. (a, d) Theoretically calculated magnon spatial distribution for different $\Omega _S$. $x\%$ is the storage efficiency, and OD is the optical depth of the atomic ensemble. (b, e) Theoretically calculated DSP spatial distribution for different $\Omega _S$. $I$ is the temporal envelope overlap ratio. Here, $\Omega _{S}$ is the same as in (a, d) with corresponding colors. Solid line: magnon input. Dashed line: photon input. (c, f) Experimentally measured ${g^{(2)}(\Delta\tau=0,\phi _{rt}=0)}$ with different $\Omega _{s}$. $z/L_{0}$ is the normalized length of the atomic ensemble and $\gamma_{31}=2\pi\times3 MHz$ is the dephasing rate between $|3\rangle$ and $|1\rangle$. The error bars in (c) and (f) indicate $1\sigma$ standard error from five measurements. (b) and (c) are data for OD = 30, and (e) and (f) are data for OD = 150.}
\end{figure*}

As shown in Fig.~\ref{Fig1}(c), our experimental setup contains two magneto-optical traps of $^{85}$Rb atoms (MOT$_1$, MOT$_2$), which serve as a heralded single photon source and  an NHBS based on quantum memory, respectively. In MOT$_{1}$, narrowband photon pairs with a Gaussian temporal shape are generated via a spontaneous four-wave-mixing process~\cite{LiJF2019,Du2015PRL,Yan2014PRL}. First, the laser-cooled atoms are optically pumped to the lowest hyperfine level $|5S_{1/2}, F=2\rangle$ at the beginning of each experimental cycle. The OD of the atomic ensemble is 120 on the transition $|5S_{1/2}, F=2\rangle\leftrightarrow|5P_{1/2}, F=3\rangle$. Then, a pair of counter-propagating pumping laser beams (780 nm, $\sigma^{-}$, 14 $\mu W$) and coupling laser beams (795 nm, $\sigma^{+}$,  3 $mW$ ) are shone on MOT$_{1}$ with an angle of $2.75^{o}$ to the quantization axis. A counter-propagating entangled photon pair consisting of a Stokes photon and an anti-Stokes photon ($\omega_{Si}$ and $\omega_{ASi}$) are collected along the quantization axis. The full width at half maximum of the temporal waveform is about 100\,ns. The conditional second order self-correlation function $g_e^{(2)}$ for $\omega_{AS}$ is measured to be $0.22\pm0.03$ with a Hanbury-Brown-Twiss interferometer~\cite{HBT1986}, which indicates an excellent single-photon nature.

MOT$_{2}$ is realized with a typical EIT-based quantum memory setup~\cite{Yan2019Qmem}. Similar to MOT$_{1}$ , the laser-cooled atoms are initialized to the Zeeman state $|5{S_{1/2}},F = 2,{m_F} = 2\rangle$ in the experiments. The OD of the atomic ensemble can be controlled from 30 to 150 on demand. The control laser is shone on the ensemble with an angle of $1^{o}$ and a beam waist of 2.3 mm. By manipulating the DSP through the control laser duration, we could realize  not only the hybrid BS, but also the well-performed reversible quantum memories~\cite{Yan2019Qmem,Yu2018PRL,Laurat2018NC,KSu2021}. Therefore, a single magnon is prepared by storing a single photon in an EIT-based quantum memory. The magnon can then be detected after being converted back into a photon. The measured conditional second order self-correlation function $g_m^{(2)}$ for the magnon prepared in MOT$_{2}$ is $0.27\pm0.06$, which  shows good single particle nature.

The quantum interference experiment runs periodically with a repetition rate of 150 Hz. In each cycle, the experimental time window following the MOT loading is 0.5 ms. When a single magnon is prepared, another single photon $\omega_{ASi+1}$ produced by MOT$_1$ arrives at  MOT$_{2}$, and the quantum interference between the magnon and photon is implemented by switching on the control laser ($\Omega _{BS}$). During this process, there exist two overlapping DSPs inside the medium. When part of the DSP leaves the EIT medium back to photons (the photon output port), the control laser ($\Omega _{BS}$) is switched off and the rest  of the DSP is converted back to pure magnons ( the magnon output port). The outputs are measured separately:  the photonic output port is detected by a single photon counting module while the magnonic output is detected by switching on the control laser ($\Omega _{R}$) and converting the magnons to photons. The second order  magnon-photon cross-correlation  after the HOM interferometer can thus be measured (Sec. IV of the Supplemental Material~\cite{SM}):
\begin{equation}
\begin{aligned}
 &{g^{(2)}}(\Delta {\tau},{\phi _{rt}}) = \left( {1 + I\left( {\Delta {\tau}} \right)\cos \left( {{\phi _{rt}}} \right)} \right),
\end{aligned}
\end{equation}
where $\Delta {\tau}$ represents the time difference between the magnon and the photon and $I\left( {\Delta {\tau}}\right)$ represents the temporal envelope overlap ratio.  The time difference $\Delta \tau$  is controllable in our experiments by manipulating the parameter $\Omega_{BS}$. For ${g^{(2)}}=2$ or $0$ and ${\phi _{rt}}=0$ or $\pi$, the NHBS dominates by the non-Hermitian or Hermitian nature, respectively. For ${g^{(2)}}=2$, we have the $|1,1\rangle$ magnon-photon output state, which indicates a Fermion-like statistic interaction between two distinct bosons.

\begin{figure}
\centering\includegraphics[width=8.5cm]{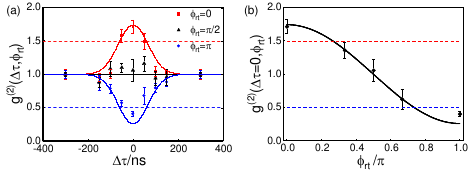}\\
\caption{\label{Fig3} Hermiticity of the magnon-photon interference.
(a) The measured ${g^{(2)}}$ with different $\Delta\tau$. Red square: ${\phi _{rt}}=0$ with $OD=30$ and $\Delta=0 ~MHz$. Blue circle: ${\phi _{rt}}=\pi/2$ with $OD=66$ and $\Delta=30 ~MHz$. Black triangle: ${\phi _{rt}}=\pi$ with $OD=100$ and $\Delta=60 ~MHz$. The solid lines are the theoretically fitted results. (b) The measured ${g^{(2)}}$ with different ${\phi _{rt}}$. The solid line is the theoretically fitted result. The dashed lines are the classical limit values. The error bars represent the $1\sigma$ standard error from five measurements.}
\end{figure}

\begin{figure*}
\centering\includegraphics[width=16cm]{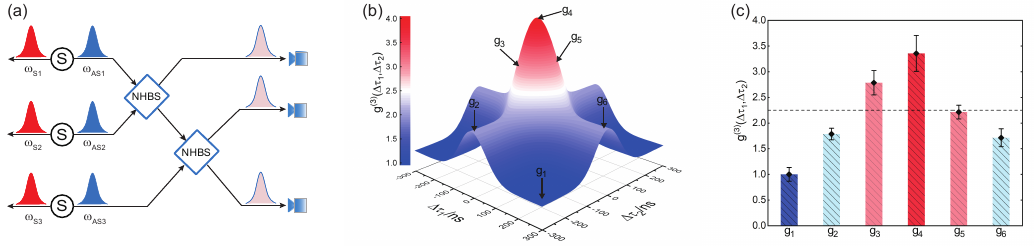}\\
\caption{\label{Fig4} Three-photon interference.
(a) Experimental scheme. $S$ is the paired photon source ($\omega_{Si}$ and $\omega_{ASi}$) and NHBS is the quantum memory based non-Hermitian beam splitter. (b) The theoretically calculated $g^{(3)}(\Delta\tau_{1},\Delta\tau_{2})$ with different time delays $\Delta\tau_{1}$ and $\Delta\tau_{2}$. (c) The experimental results of $g^{(3)}(\Delta\tau_{1},\Delta\tau_{2})$, corresponding to the points signed in (b). The dashed line is the classical limit value. The error bars in (c) indicate $1\sigma$ standard error from five measurements.}
\end{figure*}

For an ideal HOM interference, $I\left( {\Delta {\tau}}\right)=1$ is expected when $\Delta {\tau}=0$. However, completely overlapping two DSPs in the EIT medium is challenging. In our work, we could maximize the overlap by tuning $\Omega _S$ to optimize the interference contrast. The spatial distribution of the magnon over the atomic medium is calculated in the storage process, as shown in Fig. \ref{Fig2}(a) and (d). We also numerically simulate the evolution of both DSPs when another photon enters the EIT medium. As shown in Fig. \ref{Fig2}(b) and (e), the overlap ratio described by the normalized overlap integration between two DSPs is determined by the initial position of the prepared magnon. The above theoretical results are calculated through the optical Bloch equations (Eq.S6 in Supplemental Material \cite{SM}). The maximum overlap ratio can be obtained by choosing a proper $\Omega _S$ in the storage process. The splitting ratio $|t_{1}/r_{1}|(|t_{2}/r_{2}|)$ is controlled by the switching off time of the $\Omega _{BS}$. In the experiment, $\Omega _{BS}$ is switched off at the centre of the two DSPs, thus $|t_{1}/r_{1}| \approx |r_{2}/t_{2}|$. As shown in Fig. \ref{Fig2}(c) and (f), the second order cross-correlation $g^{(2)}$ is measured with different $\Omega _S$ and different ODs. High interference contrast can be realized with different ODs by controlling $\Omega _S$. The maximum $g^{(2)} = 1.75\pm 0.09$ is realized in the experiment, which is limited by the maximum achievable  overlap ratio.

With $|t_{1}|^{2}=0.15$,$|r_{1}|^{2}=0.20$, $|t_{1}|^{2}+|r_{1}|^{2}\approx0.35$, $|t_{2}|^{2}=0.26$,$|r_{2}|^{2}=0.22$, and $|t_{2}|^{2}+|r_{2}|^{2}\approx0.48$, we observed the optimal interference contrast. The loss of the magnon is determined by $\Omega _S$ in the storage process. The loss of the photon is determined by $\Omega _{BS}$ in the interference process. Under these conditions, the non-Hermicity of the BS can be further tuned by controlling ${\phi _{rt}}$ from $0$ to $\pi$ according to Eq.(2). With an acousto-optic modulator, the detuning $\Delta$ of the control laser can be arbitrarily controlled and thus change the relative phase ${\phi _{rt}}$ from $0$ to $\pi$, which indicates the transition from a non-Hermitian BS to a Hermitian BS. This relative phase usually shows no significant effect on single particle interference but fundamentally affects  two-particle interference. Here, we experimentally demonstrate that, by tuning the non-Hermicity of the BS, we can transition the bosonic HOM interference to the fermionic HOM interference between two bosons. As shown in Fig. \ref{Fig3}, the HOM interference bump and dip are observed in the experiment with different $\Delta$. The maximum ${g^{(2)}}=1.71$ of the bump and the minimum ${g^{(2)}}=0.4$ of the dip are measured, which significantly violate the respective classical limit of the second order cross-correlation of 0.5 and 1.5 \cite{SM,Chenjfcpl2020}. In our experiment,  loss is kept stable by adjusting the resonant OD for different $\Delta$. Our results agree well with  theoretical simulations and  indicate that a quantum two-particle interference is realized.

We next  utilize our NHBS to realize three-photon interference. When ${\phi _{rt}}=0$, the third-order cross-correlation of  three output photons can be described~\cite{SM} by
\begin{equation}
 \begin{aligned}
  &{g^{(3)}}(\Delta {\tau_{1}},\Delta {\tau_{2}}) =\left( {1 + I\left( {\Delta {\tau_1}} \right)} \right)\left( {1 + I\left( {\Delta {\tau_2}} \right)} \right),
 \end{aligned}
\end{equation}
where $\Delta\tau_{1}$ and $\Delta\tau_{2}$ represent the time difference between  the first-second and second-third photons, respectively. Three photons sequentially generated from a single photon source are guided one by one  into the quantum memory based on NHBS, as shown in Fig.~\ref{Fig4}(a). With ${\phi _{rt}}=0$, three-photon anti-bunching effects can be observed in three output ports [Fig.~\ref{Fig4}(b)]. In an ideal three-photon NHBS interference, the third order cross-correlation function $g^{(3)}$ of photons in output ports is expected to be $4$ while the classical threshold is $1.5^{2},$ see Supplemental Materials~\cite{SM}. As shown in Fig.~\ref{Fig4}(c), we achieve a maximal $g^{(3)}$ of $3.36\pm 0.35$ with $\Delta\tau_{1}=\Delta\tau_{2}=0$, which shows that  multi-particle interference works in the quantum regime. In this interference scheme, a post-selected tri-photon state of $|1,1,1\rangle$ can be produced in a heralded manner, and it shows a  fermion-like statistic behavior.

In conclusion, we have demonstrated quantum interference between non-identical bosonic particles (a single-magnon and a single-photon) using an atomic hybrid BS with tunable splitting ratio, loss, and non-Hermicity. In our magnon-photon quantum interferometer, both bosonic bunching and fermionic anti-bunching are observed by controlling the non-Hermicity of the hybrid magnon-photon beam splitter. Furthermore, multi-particle interference that simulates the behavior of three fermions by three input photons is realized. Our result validates and deepens the  understanding of quantum interference and could stimulate further experimental and theoretical investigations of the quantum effects between distinction bosons or fermions. Our work also unveils  the cold atom quantum memory as a versatile platform, which has the advantages of a long coherence time, multiple polariton modes, and fully controllable parameters of the generalized beam splitter. The cold atom quantum memory has potential  for studying fundamental quantum mechanics and realizing innovative quantum devices for linear optical quantum computation, boson sampling, and quantum walk~\cite{PJW2020Qadv,QW2012,QW2013SR}.

\bigskip
\begin{acknowledgments}
We thank Prof. Jiefei Chen for helpful discussions.
This work was supported by the National Key Research and Development Program of China (Grant No. 2020YFA0309500), the Key-Area Research and Development Program of Guangdong Province (Grant No. 2019B030330001 and No. 2020A1515110848), and the National Natural Science Foundation of China (NSFC) through Grants (No. 62005082, No. 12004120, No. U20A2074, and No. U1801661), the Natural Science Foundation of Guangdong Province (Grant No. 2018A0303130066), the China Postdoctoral Science Foundation (Grant No. 2020M672681, No. 2021M691102). C.-L.Z. was supported by  NSFC Grants (No. 11922411 and No. 11874342).

K. Su, and Y. Zhong contributed equally.

\end{acknowledgments}

\end{document}